\documentclass[twocolumn,aps,showpacs,amssymb]{revtex4}

\usepackage{graphicx} 

\usepackage{ulem}

\begin{document}


\title{Evaporation of buffer gas-thermalized anions out of a multipole rf ion trap}

\author{J. Mikosch}

\author{U. Fr\"uhling}

\author{S. Trippel}

\author{D. Schwalm} \altaffiliation{Max-Planck-Institut f{\"u}r Kernphysik,
Saupfercheckweg 1, 69117 Heidelberg, Germany; Present affiliation: Joseph
Meyerhoff Visiting Professor, Weizmann Institute, Rehovot, Israel}

\author{M. Weidem\"uller}

\author{R. Wester} 
\email{roland.wester@physik.uni-freiburg.de}

\affiliation{Physikalisches Institut, Universit{\"a}t Freiburg,
Hermann-Herder-Stra{\ss}e 3, 79104 Freiburg, Germany}

\date{\today}

\begin{abstract}
We identify plain evaporation of ions as the fundamental loss mechanism out of
a multipole ion trap. Using thermalized negative Cl$^-$ ions we find that the
evaporative loss rate is proportional to a Boltzmann factor. This
thermodynamic description sheds new light on the dynamics of particles in
time-varying confining potentials. It specifically allows us to extract the
effective depth of the ion trap as the activation energy for evaporation. As a
function of the rf amplitude we find two distinct regimes related to the
stability of motion of the trapped ions. For low amplitudes the entire trap
allows for stable motion and the trap depth increases with the rf field. For
larger rf amplitudes, however, rapid energy transfer from the field to the ion
motion can occur at large trap radii, which leads to a reduction of the
effective trapping volume. In this regime the trap depth decreases again with
increasing rf amplitude. We give an analytical parameterization of the trap
depth for various multipole traps that allows predictions of the most
favorable trapping conditions.
\end{abstract}


\maketitle


Evaporation of atoms and molecules out of a confined thermalized ensemble is a
well understood process \cite{davis95a,luiten96} and represents the decisive
cooling step towards Bose-Einstein condensation \cite{petrich95,davis95b}. In
radiofrequency (rf) ion traps, early studies of the dynamics have shown that,
in the absense of a buffer gas, thermalization is acchieved by ion-ion
collisions and evaporative losses \cite{dehmelt67,church88}. In applications
of rf Paul traps \cite{paul90} for quantum information processing
\cite{haffner05,reichle06}, precision spectroscopy
\cite{oskay06,schneider05,jelenkovic06}, and the production of translationally
ultracold molecular ions \cite{drewsen04,blythe05}, trapped ions are cooled
using light forces. For the sympathetic cooling of ions in a buffer gas,
higher order multipole rf traps at cryogenic temperature have proven to be
more useful due to their large field free region
\cite{gerlich95,schlemmer99}. Multipole ion traps are hence widely used to
prepare cold molecular ions in laboratory astrophysics \cite{gerlich06}, to
reduce the Doppler shift in microwave ion clocks \cite{prestage01}, to measure
absolute photodetachment cross sections of molecular anions \cite{trippel06},
the photofragmentation of biomolecules \cite{boyarkin06}, for precision
rovibrational spectroscopy \cite{asmis03,mikosch04,asvany05} and for collision
experiments \cite{kreckel05}.

If the ion temperature is fixed by collisions with a buffer gas, evaporative
loss measurements provides access to the stability of ion motion in traps. In
a quadrupole or Paul trap, stable ion motion is described analytically by the
Matthieu equations, which puts well-defined constraints on the trapping
fields. In addition to these global boundaries of stability, heating of
trapped ions by the rf field might be assisted or damped by Coulomb
interaction, and phase transitions between crystalline and chaotically moving
clouds are observed \cite{bluemel88}. In contrast, for ions moving in a high
order multipole field the equations of motion have no analytical
solution. There have been attempts to numerically establish a stability
diagram for two-dimensional hexapoles and octupoles \cite{haegg86a}. For
different starting conditions the obtained stability diagrams look very
different and the regions of stability do not show well defined boundaries
opposite to the case of the quadrupole trap.

In this letter we show that one can attribute in a thermodynamic sense an
effective trap depth to an ion cloud in a multipole ion trap. For this purpose
we measure the rate of evaporation of trapped ions as a function of their
translational temperature controlled by a bath of helium buffer gas. We use
Cl$^-$ anions with a high electron affinity, because one can safely ignore
losses due to parasitic chemical reactions with the background gas. The
measured evaporation rate is found to be proportional to a Boltzmann factor
with the activation energy given by an effective trap depth, i.e. the minimal
kinetic energy ions need to escape from the trap. We study this effective trap
depth as a function of the applied rf amplitude and find two distinct
regimes. Based on numerical calculations, the steeply rising trap depth for
low rf amplitudes is explained by an entirely adiabatic motion of the trapped
ions, whereas the regime of almost constant trap depth for high rf amplitude
shows the appearance of regions of unstable ion motion inside the trap. An
analytical model allows us to predict the parameters for a maximum trapping
potential and trapping volume.

\begin{figure}[tb]
  \center \includegraphics[width=0.9\columnwidth]{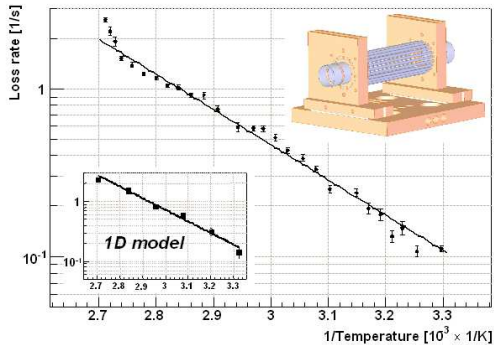}
  \caption{Measured loss rate of Cl$^{-}$ anions out of the 22pole trap (see
  schematic view) versus temperature at an rf amplitude of $V_0=33$\,V. The
  solid line is a fit to Eq.\ (\ref{arrhenius_eqn}). A similar result is
  obtained with a 1D numerical model (see inset) in the same temperature range
  at an rf amplitude of $V_0=10.5$\,V.}
  \label{fig1}
\end{figure}

The 22pole ion trap used in our experiment \cite{trippel06} approximates a
multipole rf field of order $n=11$. Storage is achieved by 22 stainless steel
rods (1\,mm diameter) forming a 40\,mm long cylindrical cage (inscribed
diameter 2\,$r_0$\,=\,10\,mm, see schematic view in Fig.\ \ref{fig1}). The
rods are alternatingly connected to the two ports of an rf oscillator ($\omega
= 2\pi\times 4.7\,$MHz) providing a cylindical effective potential in the
radial direction. Along the axis ions are confined by small dc voltages
(3-10\,V) applied to cylindrical entrance and exit electrodes. We manipulate
the temperature of the ion ensemble via heating or cooling the trap and its
housing, into which we apply helium at a well defined density of typically
$2\times10^{14}$\,cm$^{-3}$. Thermalization of the ions hence occurs on the
timescale of 100\,$\mu$s, assuming a Langevin-limited collision rate. The ion
source employs a pulsed supersonic expansion of argon with a small admixture
of CCl$_4$ bombarded by a pulsed 1\,keV electron beam. In the created local
plasma Cl$^-$ ions are formed via dissociative attachment of slow electrons to
CCl$_4$ and transferred to the trap with a time-of-flight Wiley-McLaren mass
spectrometer. To optimize the trapping efficiency ions are initially trapped
at a fixed rf amplitude for 200\,ms, which is then ramped down linearly within
100\,ms to the desired value. Typically 2$\times10^3$ Cl$^-$ anions are
trapped per filling, which results in a Spitzer self collision time
\cite{church88} of about 40\,s at 300\,K, ensuring that ion-ion interactions
do not play a role. After extraction the ions are mass analyzed in a second
time-of-flight stage before being detected on a microchannel plate.

We observe that the loss rate of Cl$^{-}$ anions from the trap depends
strongly on their temperature as can be seen from Fig.\ \ref{fig1}. For this
measurement we heat the trap to 370\,K and measure the loss rate while slowly
cooling the trap to 300\,K. Each loss rate is derived by measuring the decay
of the chlorine anion peak in the mass spectrum when increasing the storage
time. It is important to note that we do not observe any other peaks in the
mass spectra for any of the storage times. Actually, chemical reactions with
any of the possible impurities in the background gas are not expected
considering the high electron affinity of chlorine of 3.6\,eV. Destruction of
the anions through detachment of the excess electron by blackbody radiation is
excluded on this timescale regarding the high photon energies
needed. Similarly, detachment by collisions can be excluded in the temperature
regime under study. This indicates, that the ions are physically removed from
the trap and that the only possible loss process from the trap is evaporation
over the effective energy barrier formed by the rf field. This is strongly
supported by the dependence of the loss rate on the temperature of the ion
ensemble. Fig.\ \ref{fig1} reflects that by increasing its temperature we
force more ions over the barrier.

\begin{figure}[tb]
  \center
  \includegraphics[width=0.8\columnwidth]{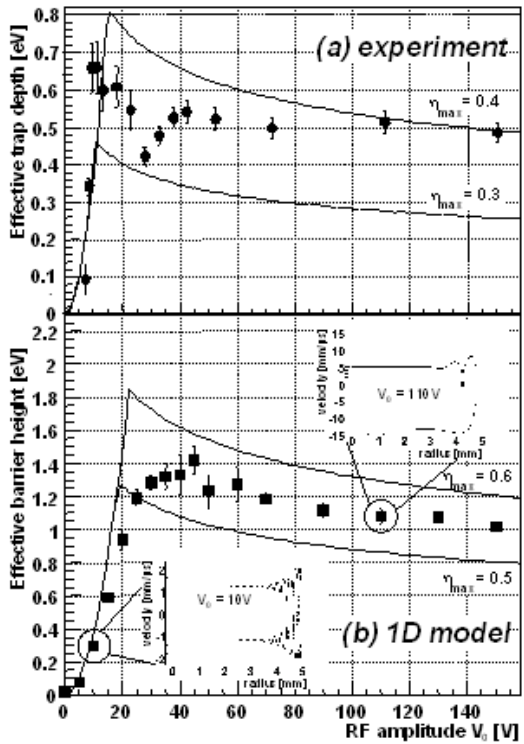}
  \caption{Experimentally determined effective trap depth (a) and numerically
    obtained effective barrier height (b) as a function of the rf amplitude.
    Each point is derived from a fit similar to Fig.\ \ref{fig1}. Inserted
    lines stem from the analytical model for different parameters of the
    maximum adiabaticity parameter $\eta_{max}$. The two inserted phase space
    trajectories represent ions with a starting velocity just below the
    threshold for direct escape. At low rf amplitudes the ion returns from the
    barrier with the initial velocity. Strong transfer of energy from the
    field to the ion motion is seen at high rf amplitudes.}
  \label{fig2}
\end{figure}

We fit the measured loss rate $k(T)$ to a Boltzmann factor
\begin{equation}
\label{arrhenius_eqn}
k(T) = A \times e^{-\frac{E_a}{k_B\,T} },
\end{equation} 
with the activation energy for loss of anions $E_a$ and the pre-exponential
factor $A$ as free parameters. $E_a$ is directly obtained in units of the
Boltzmann constant $k_B$ as the slope of the fit in Fig.\ \ref{fig1}. $E_a$
represents the kinetic energy ions need to exceed the effective barrier formed
by the rf field and therefore provides an elegant way to determine the
effective trap depth. We have determined the effective trap depth of the
22pole ion trap as a function of the applied rf amplitude $V_0$. Experimental
accuracies are derived from the Boltzmann fits. As can be seen from Fig.\
\ref{fig2} the effective trap depth $U(V_0)$ rises steeply for low rf
amplitudes $V_0$ and reaches a maximum of 0.65\,eV at around 12\,V. For higher
amplitudes $U(V_0)$ decreases slightly before it eventually levels off at
0.5\,eV for amplitudes $V_0\,>\,40$V. This shows that trapping is
characterized by two distinct regimes, one with a fast increasing trap depth
and one where the depth is roughly constant. In addition a small dip in the
measured effective trap depth is observed at an rf amplitude of 30\,V.

Additional evidence for the two different evaporation regimes stems from the
pre-exponential factor $A$ in equation~(\ref{arrhenius_eqn}). For large rf
amplitudes $A$ is found to be of the order of $10^7/s$, which fits best to the
frequency of the field, and is independent of the helium density up to
$2\times10^{15}$\,cm$^{-3}$. For very small rf amplitudes $A$ is two to three
orders of magnitude smaller and compares well with the collision frequency of
the anions with the helium buffer gas.

To get a microscopic insight into the involved loss processes, we have set up
a simple numerical model; details will be published elsewhere. Cl$^-$ ions are
propagated in a one-dimensional oscillating electric multipole field of order
$n$ \cite{gerlich92} $E(r,t) = \frac{V_0}{r_0} \ n \ {\vert\,r\,\vert}^{n-1}
\cos(\omega t + \Phi)$ with $-r_0 \leq r \leq +r_0$ by solving the equation of
motion numerically.  We record the loss of ions starting at $r=0$ with random
phase $\Phi$ with an initial velocity drawn from a Maxwellian
distribution. The inset in Fig.\ref{fig1} shows the temperature dependence of
the loss rate. In analogy to the experiment we determine the height of the
effective barrier as a function of the rf amplitude. As seen in Fig.\
\ref{fig2}b the numerically determined effective barrier height rises steeply
for low rf amplitudes, reaches a maximum and then slightly decreases for high
amplitudes. This behavior is consistent with the corresponding experimental
data in Fig.\ \ref{fig2}a. In detail, however, there are distinct differences
in the maximum barrier height and its corresponding rf
amplitude. Nevertheless, the simple 1D model provides valuable insight. The
initial velocities of those ions lost during propagation demonstrate that
exclusively hot ions in the Boltzmann tail of the distribution are lost, which
backs our analysis of evaporation-driven ion loss. The calculated ion
trajectories in phase space are characterized by a uniform velocity in the
region of low field at smaller radii evolving into a fast oscillation of the
velocity in the region of high field near the turning point. For a small rf
amplitude $V_0$ where in Fig.\ref{fig2}b the effective barrier height is found
to be steeply rising, no transfer of energy from the field is observed. Either
the ions are reflected from the flapping potential, preserving their kinetic
energy, or their initial kinetic energy exceeds the effective barrier and ions
are lost on first approach. The inset in Fig.\ref{fig2}b for an rf amplitude
of $V_0$ = 10\,V shows such a trajectory for an ion starting with a velocity
just below the limit for getting lost. For high rf amplitudes, however, ions
with an initial velocity just below the limit for getting lost on the first
approach are found to rapidly aquire energy from the field. This can be seen
from the trajectory shown in Fig.\ref{fig2}b for an rf amplitude of $V_0$ =
110\,V, where the ion velocity on return is more than twice the initial
value. The ion will not survive the next approach to the barrier.

By comparison with the numerical simulation we are able to interpret the
abrupt break-off in the experimentally observed effective trap depth shown in
Fig.\ref{fig2}a and the accompanied change in the pre-factor $A$ to a value of
the order of the rf frequency. We identify it with the advent of non-adiabatic
ion motion. Ions are lost from the trap by rapidly acquiring translational
energy from the rf field within a few rf cycles, once they happen to be
located in a trap region where the rf amplitude is strong enough to disturb
adiabatic motion.

For an analytic description of the trap depth we introduce the trapping volume
as the volume in which stable trajectories without transfer of energy from the
field are possible. The numerical simulation shows in agreement with our
measurement that the trapping volume is reduced for increasing rf amplitude
and that the reduction prevails over the increase in the repelling force by
the rf field. We parameterize the trapping volume based on the adiabaticity
parameter $\eta$ defined in \cite{gerlich92}, which scales with the
$(n-2)$-power of the radial position of the ion. We consider the trapping
volume to be bound by a critical radius $r_{crit}$ where $\eta = \eta_{max}$
if this radius is smaller than the geometrical radius $r_0$ of the trap,
otherwise by $r_0$. We then identify the time-independent effective potential
at the edge of the trapping volume with the effective trap depth. This results
in an effective trap depth of an ideal multipole trap of
\begin{equation}
\label{eqn:sim:trapdepth}
U\,(V_0) = \frac{1}{8} \ \frac{(q V_0)^2}{\epsilon} \ r_{tv}^{2n-2}
\end{equation} 
where $r_{tv} = {\mathrm Min} \left\{r_0,\,r_{crit}\right\}$ is the radius of
the trapping volume, $r_{crit}$ the critical radius
\begin{equation}
r_{crit} = r_0 \ \left(\eta_{max} ~ \frac{n}{n-1} ~ \frac{\epsilon}{qV_0}\right)^{1/(n-2)}, 
\end{equation} 
and $\epsilon = 1\,/\,(2n^2)\,m\omega^2r_0^2$ the characteristic energy. For
low rf amplitudes $V_0$ the adiabaticity parameter $\eta$ does not exceed
$\eta_{max}$ in the entire trap. Here the trapping volume equals the
geometrical volume of the trap and the effective trap depth is given by the
effective potential at radius $r_0$. In this regime, the effective trap depth
grows quadratically with the applied rf amplitude $V_0$.  The maximal trap
depth is obtained, when $\eta$ reaches the largest allowed value $\eta_{max}$
just at the geometrical edge of the trap. For higher amplitudes, the region of
non-adiabatic ion motion begins to penetrate into the trap. The trapping
volume is now bound by the condition $\eta\,=\,\eta_{max}$ and the effective
trap depth is given by the effective potential at this reduced radius, which
scales as $V_0^{-2/9}$.

The lines in Fig.\ref{fig2}a represent the effective trap depth $U(V_0)$
according to the described analytical model for $\eta_{max}$ between 0.3 and
0.4. With these values the experimental data are very well described. Also the
numerical result in Fig.\ref{fig2}b is described well by
Eq.~(\ref{eqn:sim:trapdepth}), however with a higher value of the maximal
adiabaticity parameter $\eta_{max}$ of between 0.5 and 0.6. This difference
might be explained by weak fringe fields disturbing the perfect multipole
configuration which could originate from charge up of insulators and residual
gas deposited onto the electrodes. This is supported by the experience that
multipole ion traps provide longer lifetimes immediately after cooling down as
compared to many days of operation at low temperatures. The dip in the
effective trap depth observed at amplitudes around 25\,V is currently
unexplained and subject to further work.

\begin{figure}[tb]
\center
\includegraphics[width=0.8\columnwidth]{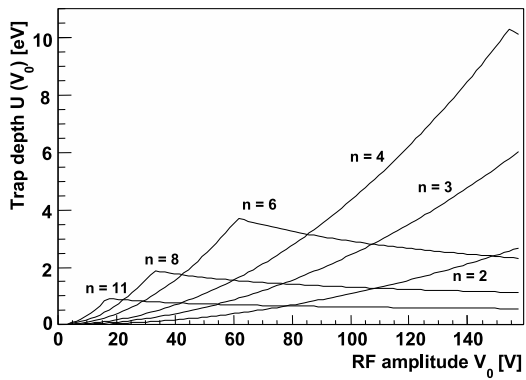}
\caption{Effective trap depths for multipoles of different order $n$ as
  derived from the analytical model assuming a maximum adiabaticity parameter
  $\eta_{max}\,=\,0.4$ (calculation for Cl$^-$ ions, r$_0=5$\,mm,
  $\omega=2\pi\times 5$\,MHz).}
\label{fig3}
\end{figure}

The analytical and numerical models can be readily applied to multipoles of
different order $n$. In Fig.\ref{fig3} the effective trap depth for Cl$^-$
ions is plotted as a function of the rf amplitude for various multipoles of
the same inscribed radius $r_0=5$\,mm operated at a fixed frequency
$\omega=2\pi\times 5$\,MHz assuming a maximum adiabaticity parameter
$\eta_{max}=0.4$. The larger field-free region for increasing multipole order
$n$ is payed by a smaller maximal effective trap depth. Also, lower order
multipole traps can be operated in an all-adiabatic mode up to higher rf
amplitudes. For a quadrupole trap ($n=2$) the adiabaticity parameter $\eta$
becomes independent of the radius, but is still a function of the rf amplitude
V$_0$. Note that this results in a distinct difference to higher order
multipoles: The trap depth scales quadratically with the rf amplitude V$_0$
until the maximal stability parameter is reached; at this point there is a
transition from adiabatic to nonadiabatic ion motion in the entire trap and
the trap depth vanishes completely.

In conclusion, we demonstrate evaporation of anions out of a multipole rf ion
trap. Analysis of the temperature-dependent evaporation rate allows us to
extract the effective trap depth. This shows that Boltzmann statistics are not
only applicable to the translational and rotational degrees of freedom of the
bulk of trapped ions \cite{schlemmer99}, but also to the high-energy tail of
the Boltzmann distribution. Furthermore, we observe the transition from an
all-adiabatic trapping to energy transfer from the field to the ion
motion. Based on a numerical calculation we introduce the concept of a
trapping volume, and by its parameterization we obtain an analytic expression
for the effective trap depth and its scaling with the rf amplitude, which is
directly applicable to multipole fields of arbitrary oder $n$.

The present results suggest possible applications of trap losses as a probing
scheme for inelastic collision processes of trapped ions, similar to schemes
used for neutral atoms \cite{staanum2006,zahzam2006}. Maximizing the trapping
volume of multipole traps based on our findings should allow one to trap an
optimum number of ions at a given space charge interaction, which is
particularly interesting for the loading of shallow surface traps
\cite{cetina07}. Furthermore, ion-ion induced evaporation in a collision free
environment may be investigated, for which we have first experimental evidence
at larger ion ensembles.

This work is supported by the Deutsche Forschungsgemeinschaft under grant
No. WE 2661/4-1 and by the Elitef{\"o}rderprogramm der Landesstiftung
Baden-W{\"u}rttemberg.

\end{document}